\documentclass[twocolumn,showpacs,preprintnumbers,amsmath,amssymb]{revtex4}

\usepackage{mathrsfs}
\usepackage{graphicx}
\usepackage{dcolumn}
\usepackage{bm}

\begin{document}

\preprint{APS/123-QED}

\title{Quantitative testing of robustness on super-omniphobic surfaces by drop impact} 

\author{Thi Phuong Nhung Nguyen$^{1,2}$, Philippe Brunet$^{2,*}$, Yannick Coffinier$^{1}$ and Rabah Boukherroub$^{1}$}
\affiliation{$^1$Institut de Recherche Interdisciplinaire (IRI) USR CNRS 3078, Universit\'e Lille Nord de France - Parc de la Haute Borne 50 Avenue de Halley, BP 70478, 59658 Villeneuve d'Ascq Cedex France, 
$^2$Institut d'Electronique de Micro\'electronique et de Nanotechnologies (IEMN) UMR CNRS 8520 - Universit\'e Lille Nord de France - Avenue Poincar\'e, BP 60069, 59652 Villeneuve d'Ascq France}


\altaffiliation{To whom correspondence should be addressed. E-mail: philippe.brunet@univ-lille1.fr}

\date{\today}

\begin{abstract}

The quality of a liquid-repellent surface is quantified by both the apparent contact angle $\theta_0$ that a sessile drop adopts on it, and the value of the liquid pressure threshold the surface can withstand without being impaled by the liquid, hence keeping a low-friction condition. We designed surfaces covered with nano-wires obtained by the vapor-liquid-solid (VLS) growth technique, that are able to repel most of the existing non-polar liquids including those of very low surface tension, as well as many polar liquids of moderate to high surface tension. These super-omniphobic surfaces exhibit apparent contact angles ranging from 125 to 160$^{\circ}$ depending on the liquid. We tested the robustness of the surfaces against impalement by carrying out drop impact experiments. Our results show how this robustness depends on the Young's contact angle $\theta_0$ related to the surface tension of the liquid, and that the orientational growth of NWs is a favorable factor for robustness.

\end{abstract}

\keywords{Superomniphobic surfaces; Drop impact; Dynamical wetting}

\maketitle

\section{Introduction}

The fabrication of water-repellent surfaces is nowadays commonly achieved by a host of different techniques \cite{BicoQuere99,Herminghaus00,Lau03,Patankar04,Gao_McCarthy06,Nosonovsky07,Verplank_etal07,Review_iemn}. It is admitted that a micro- or nano-textured surface with an appropriate low-energy coating is able to repel water. While the early studies on superhydrophobicity were undertaken on surfaces consisting of regular arrays of micron scale posts \cite{BicoQuere99,Herminghaus00}, the current state of the art benefits from recent improvements on nano-texturation by chemical growth - for instance silicon nanowires or carbon nanotubes.

However, most of the liquid-repellent surfaces are effective only for high surface-tension ($\gamma$) liquids like water. The current challenge for their applicability in lab-on-chip microfluidics or in high-performance clothes, is to design surfaces that repel liquids of lower surface tension. Recent achievements by Tuteja \textit{et al.} \cite{Tuteja07,Tuteja08} employ surfaces with overhanging roughness elements - denoted as 're-entrant', with a regular array of mushroom-shaped posts. Other kind of substrates \cite{Cao_etal08,Ramos09,Liu10} with less regular texture also achieved high repellency with low-$\gamma$ liquids.

\begin{figure}
\begin{center}
(a)\includegraphics[scale=0.32]{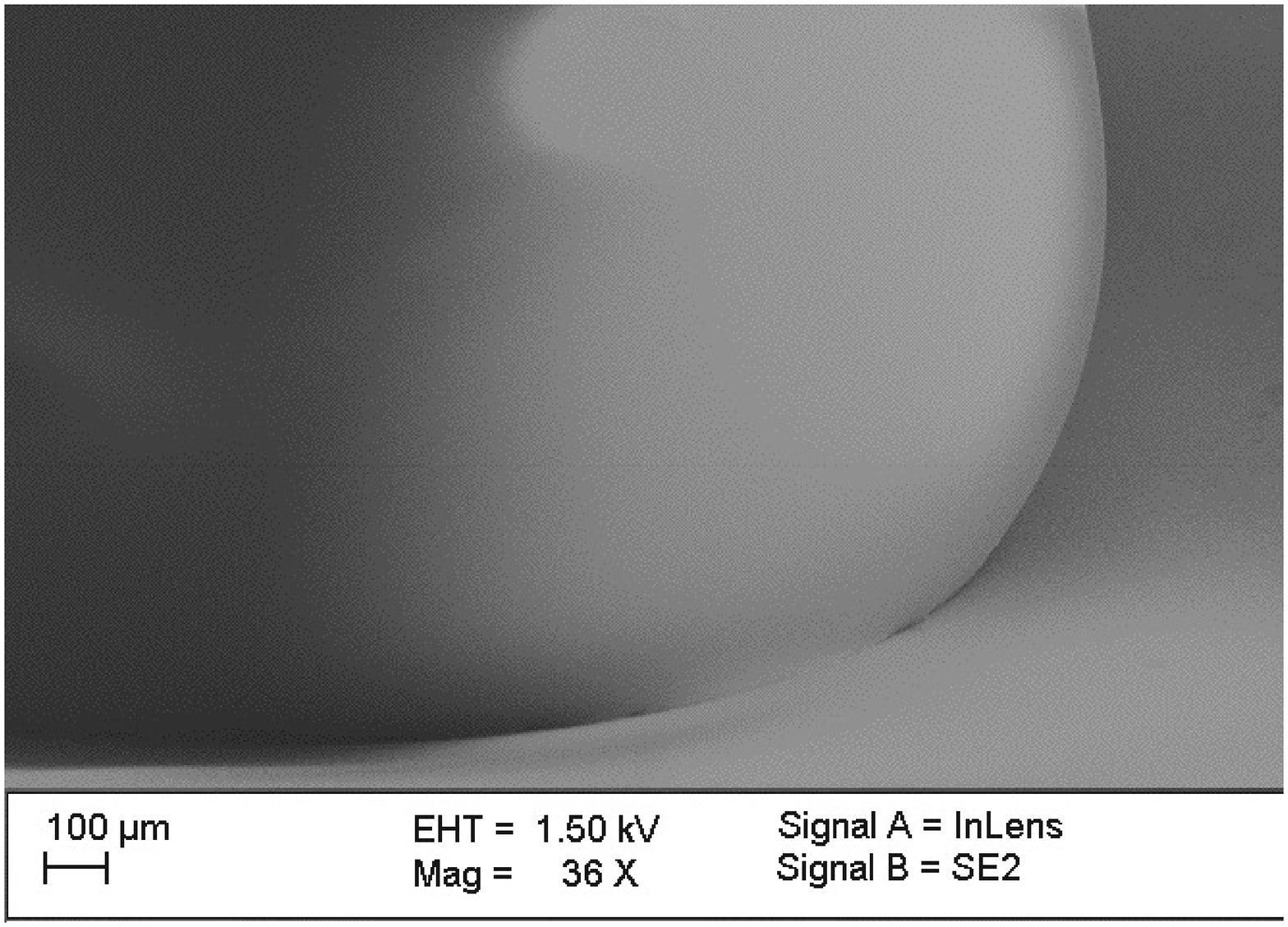}
(b)\includegraphics[scale=0.2]{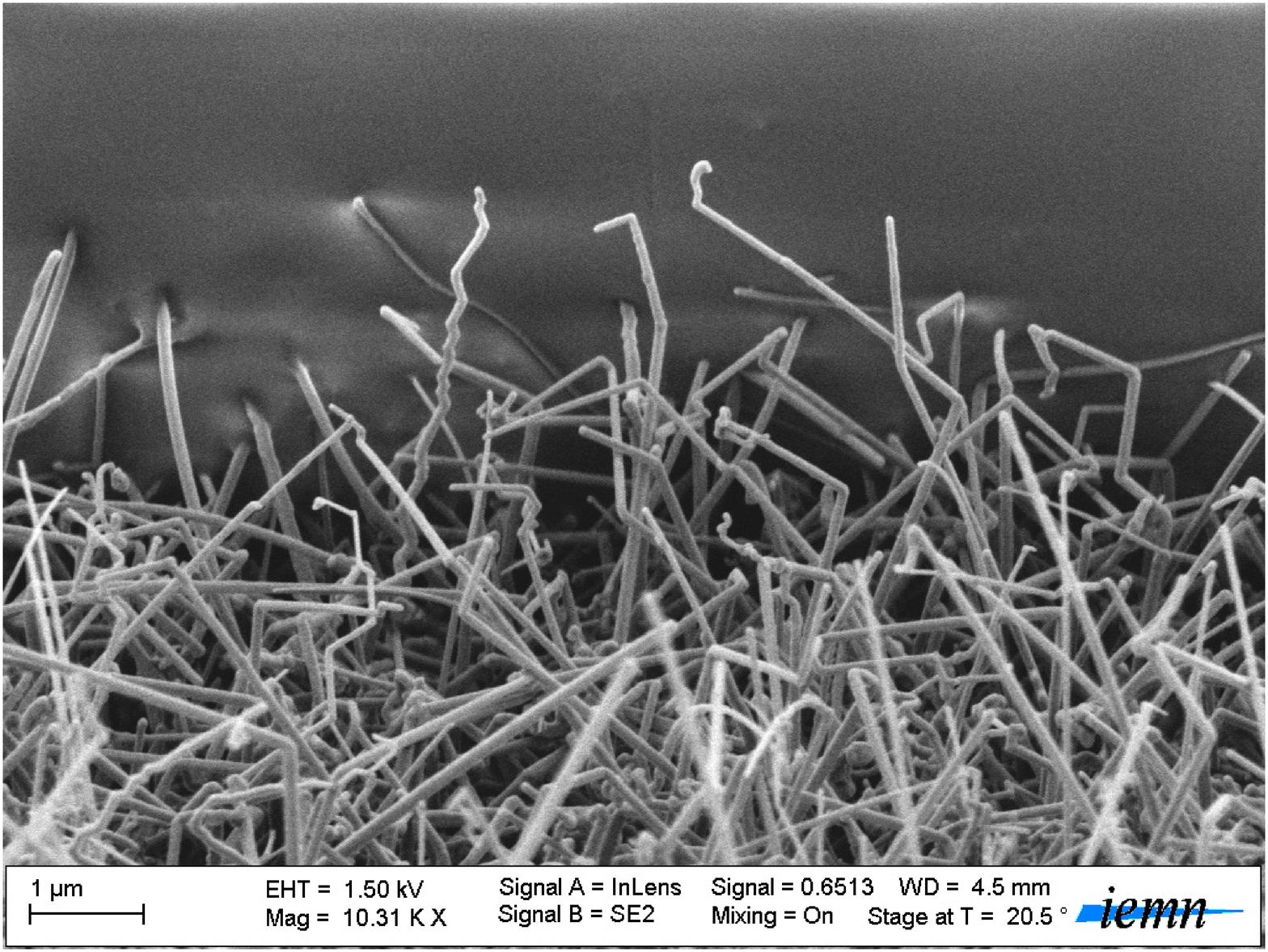}
\caption{(a) A drop of dried polymer (Polyphenyl-methylsilixane, $\gamma$=24.5 mN/cm) on a carpet of SiNWs. (b) Zoom under the drop, evidencing the Cassie-Baxter state.}
\label{fig:droppdms}
\end{center}
\end{figure}

In this paper, we present quantitative results for a new type of super-omniphobic surface: they are composed of tangled silicon nanowires (SiNWs), coated with a low surface energy Fluoro-Polymer, which fabrication details are given later on. Figure \ref{fig:droppdms} displays a scanning electron microscopy (SEM) image of a drop of Polyphenyl-methylsiloxane (Aldrich) of low surface tension ($\gamma$=24.5 mN/m) that was deposited and dried to become solid, on one of our surfaces. A contact-angle of about 120$^{\circ}$ was measured with such a usually very wetting liquid. Figure \ref{fig:droppdms}-(b) shows a magnified view close to the contact-line, which evidences the repelling character of the NWs: the drop sits on top of them. 

What defines a good super-repellent surface, is the largest possible apparent contact angle (CA) with a small contact angle hysteresis (CAH) - typically $\theta_a - \theta_r \le 5 ^{\circ}$, but also its ability to retain a state of small CAH when large external forces are applied to the liquid - thereafter, we denote 'robustness' this property. The condition for a small CAH is achieved when the liquid-vapor interface is weakly or not-impaled in the texture. Once impaled, the liquid stays stuck: the retention force and the CAH increase dramatically, which is not suitable for applications of low-friction drop displacement or self-cleaning, and this is more likely to happen for low surface-tension liquids. Therefore, one of the remaining question is how the robustness depends on the nature of the liquid ? Our study addresses this issue by measuring the impalement pressure threshold $P_c$ for liquids of various $\gamma$, using drop impact experiments.

\section{Qualitative analysis of the robustness of a surface}

It is well established that the roughness induces a modification of the \textit{apparent} macroscopic CA $\theta$. To be more precise, the Cassie-Baxter approach assumes that the drop sits on top of the roughness elements, leading to the following expression for the apparent angle \cite{Cassie_Baxter}:

\begin{equation}
\cos \theta = \phi_s (\cos \theta_0 +1) - 1 
\label{eq:cassie_baxter}
\end{equation}

\noindent where $\phi_s$ is the surface fraction of the liquid-solid interface and $\theta_0$ is the Young's contact-angle obtained on a perfect smooth surface. The Wenzel approach assumes that the liquid completely fills the space between the texture. Under this assumption, the apparent CA equals \cite{Wenzel}:

\begin{equation}
\cos \theta = r \cos \theta_0
\label{eq:wenzel}
\end{equation}

\noindent where $r$ is the surface roughness, i.e. the ratio between the total surface and the projected one ($r \ge$ 1).  Equation (\ref{eq:wenzel}) suggests that  the roughness leads to a large apparent CA, providing $\theta_0$ is larger than $\frac{\pi}{2}$, and eq.~(\ref{eq:cassie_baxter}) suggests that large $\theta$ are obtained with a texture of thin wires and a large pitch. However, eqs.~(\ref{eq:cassie_baxter}) and (\ref{eq:wenzel}) do not give any clue about which state is more favorable under given external conditions, the 'fakir' one (Cassie-Baxter) or the impaled one (Wenzel). By simple arguments on surface energy \cite{QuereReview05}, one can evaluate a critical Young's angle $\theta_{0c}$ as:

\begin{equation}
\cos \theta_{0c}  = - \frac{1- \phi_s}{r - \phi_s}
\label{eq:theta_crit}
\end{equation}

\noindent such as if $\theta_0 \ge \theta_{0c}$ the surface energy of the Cassie-Baxter state is lower than that of the Wenzel state: the drop stands on top of the texture. The condition $\theta_0 \ge \theta_{0c}$ is generally not fulfilled for low $\gamma$ liquids. However, it is possible to keep a liquid-vapor interface in a weakly impaled state even when $\theta_0 \ge \theta_{0c}$. This is related to the meta-stable character of the Cassie-Baxter state: in a free-energy landscape, this state can be located at \textit{local minimum}, whereas the free-energy is often at a \textit{global minimum} in the Wenzel state. Therefore, a perturbation of finite strength is required to jump over the energy barrier between the Cassie-Baxter and Wenzel states \cite{Nosonovsky07,Marmur08}. Furthermore, it has been recently evidenced \cite{Tuteja08,Marmur08} that a meta-stable Cassie-Baxter state can be achieved even if $\theta_0 \le \frac{\pi}{2}$, providing that the roughness elements have an overhanging shape. This peculiar geometry allowed for low-$\gamma$ liquids repellent surfaces. The robustness against impalement is then related to the height of the energy barrier between the two limit states.

Drop impact experiment is a particularly versatile way to test this robustness \cite{Bartolo06,Quere_Reyssat06,Brunet08,Tsai09,Faraday10}. During impact, the sudden vertical deceleration of the liquid particles leads to momentum transfer that applies an effective dynamical pressure $P_{dyn} = 1/2 \rho U^2$, with $U$ being the impact velocity and $\rho$ the liquid density.  The value of this pressure is easily determined by measuring $U$ just before impact, and the pressure peak is localised at the centre of impact, hence enabling the test of impalement in a very reduced area. Furthermore, it has been evidenced \cite{Bartolo06} that this dynamical pressure is equivalent to a static one for liquid impalement. In other words, there are no dynamical effects involved during the interface penetration in the texture, which ensures a universal character of the results obtained by this method.

Most of the super-omniphobic surfaces designed so far, are made of an ordered texture with roughness elements which size ranges from a few microns to a few tens of microns \cite{Tuteja07,Tuteja08}. According to the Washburn's law for pressure entry in porous media \cite{washburn}, the order of magnitude of the entry pressure is inversely proportional to the typical size $R$ of roughness: $P_w = 2 \gamma \cos \theta / R $. The quantitative measurements obtained on periodic arrays of micro-pilars evidenced that $R \sim \frac{s^2}{L}$ \cite{Bartolo06,Quere_Reyssat06}, with $s$ and $L$  stand for the space between pillars and their length, respectively. Therefore, our idea is to increase robustness by covering the surface with a forest of chemically-grown SiNWs of high aspect-ratio (from 30 to 100) and of about 50 to 100 nm in diameter, see Fig.~\ref{fig:nanofil_pure}. This texture scale is still hardly accessible to current lithography and engraving techniques: the realization of surfaces larger than about 1 cm$^2$ is expensive and time-consuming. However, this scale of texture is more simply and cheaply achievable with chemical growth of silicon NWs. To ensure an equivalent of the re-entrant geometry of surfaces used in \cite{Tuteja07,Tuteja08}, we used growth parameters (pressure, temperature, ...) in order to get disordered and multi-directional wires growing diagonally (see Fig.~\ref{fig:nanofil_pure}). The downside of this disorder, is that there does not exist obvious characterization of their roughness nor their surface fraction.

\section{Preparation of super-omniphobic substrates by chemical vapor deposition (CVD)}

The silicon NWs are synthesized on the substrate by using the vapor-liquid-solid (VLS) growth mechanism as described in more details in \cite{Verplank_etal07,Brunet08}. A 300 nm SiO$_2$ layer is deposited thermally on the silicon substrate, coated with a layer of 40 $\AA$ of gold, and then placed in an oven. The substrate is heated up to 500$^{\circ}$C, forming gold nano-particles acting as catalysts for the silicon NWs growth. Then, silane gas (SiH$_4$) is injected following its preferential decomposition on gold. The SiH$_4$ molecules dissociate and silicon atoms are incorporated into the gold droplet, forming a AuSi liquid eutectic. This induces the directional growth of a NW with the gold nano-particle on top. The NWs width, length and morphology depend on the duration, the pressure and the temperature of the process, as well as on the crystal orientation of the substrate \cite{Langmuir_flo09}. The SiO$_2$ layer ensures a relative disorder in the diagonal direction of the NWs growth. The orientation is also influenced by the pressure prescribed during the VLS process \cite{Kawashima08}: the orientation varies from about 30$^{\circ}$ to 90$^{\circ}$ with respect to the horizontal: the larger the pressure is, the straighter the NWs grow (with also a narrower distribution of orientation).

\begin{figure}
\begin{center}
(a)\includegraphics[scale=0.20]{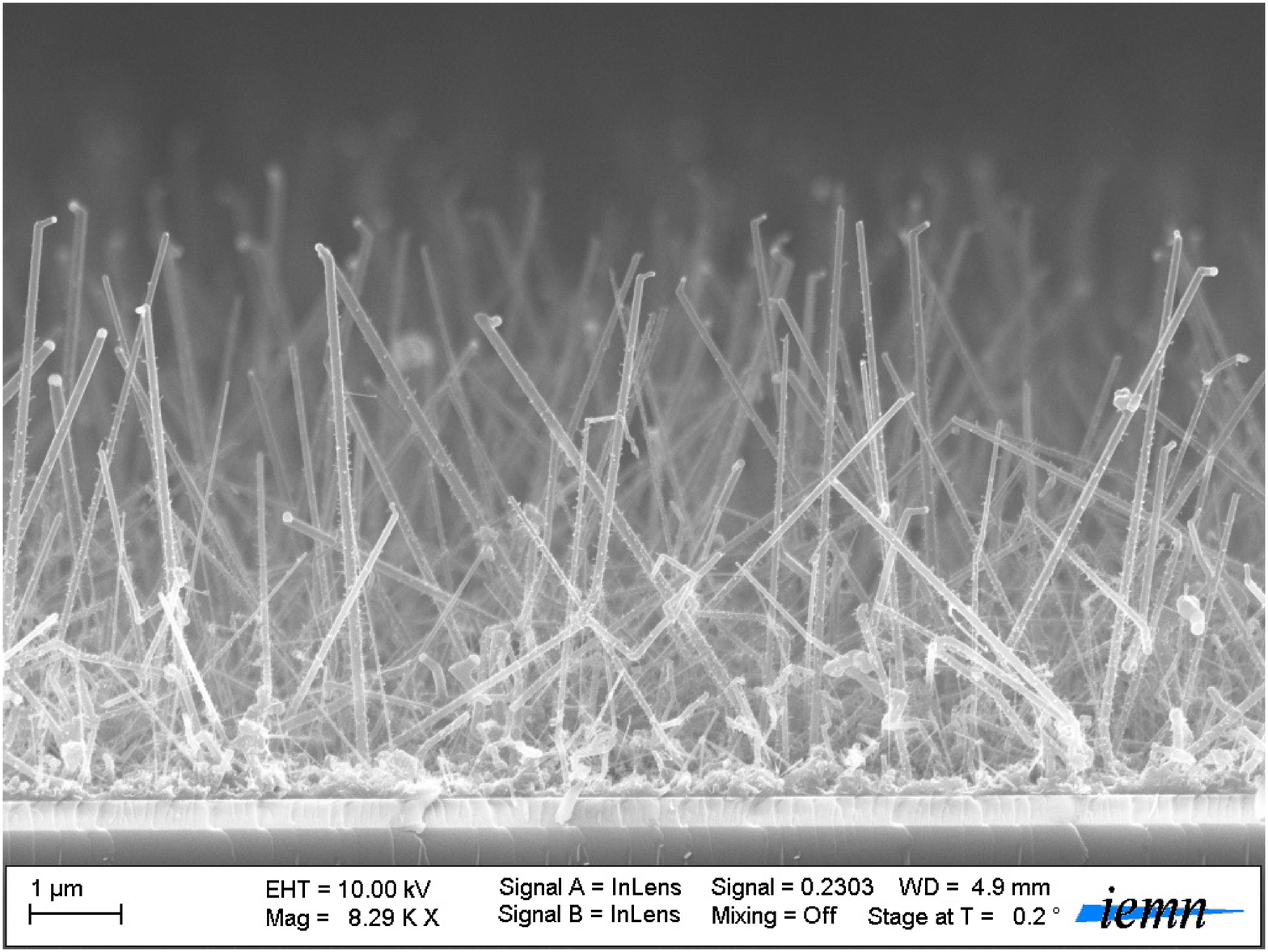}
(b)\includegraphics[scale=0.20]{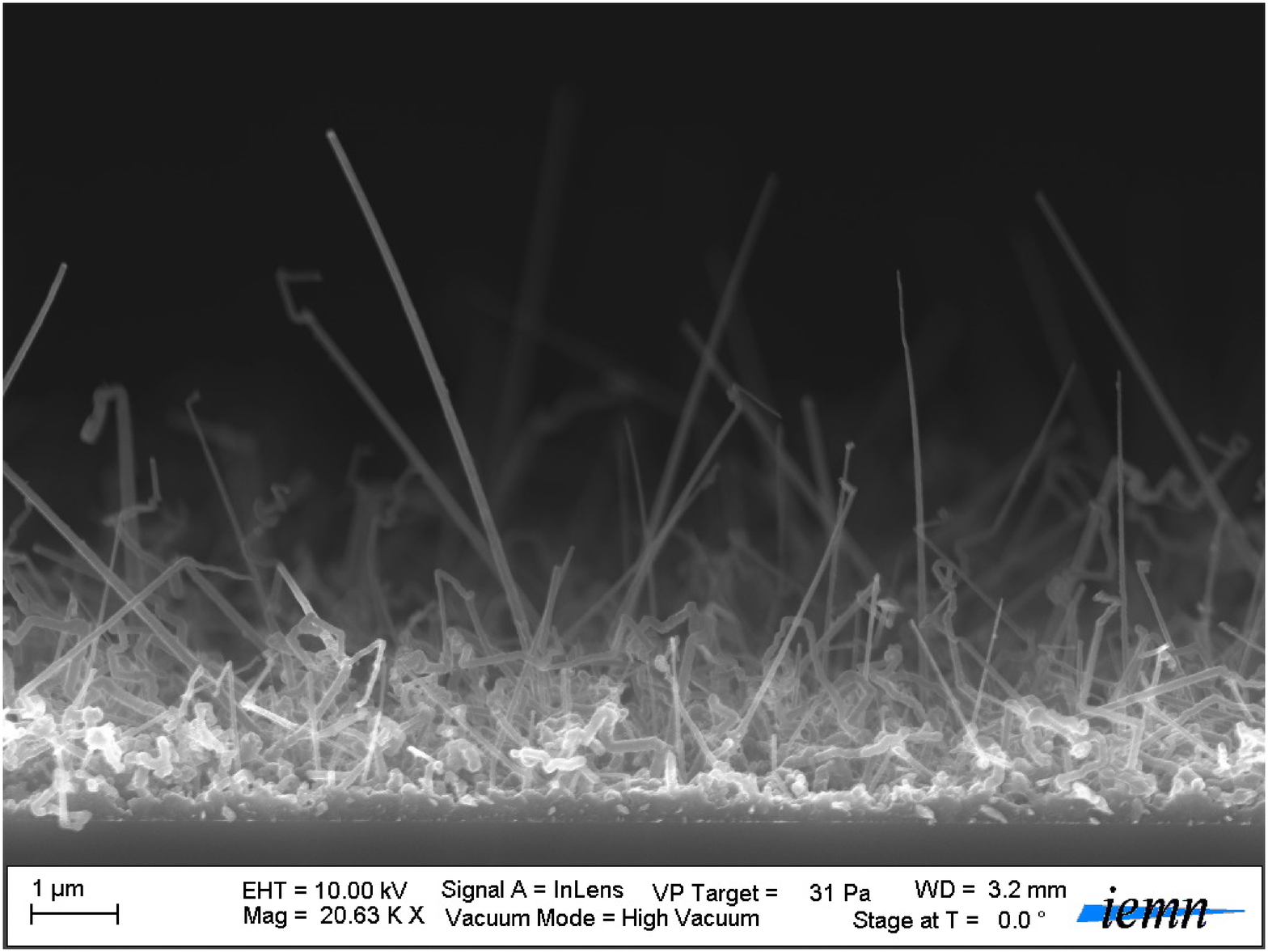}
\caption{SEM images of the NWs surfaces prepared by the VLS growth technique. (a) P=0.4 Tore and 10 min "VLS 1". (b) P=0.1 Tore and 60 min "VLS 2".}
\label{fig:nanofil_pure}
\end{center}
\end{figure}

Two surfaces have been used in this study corresponding to different growth conditions on a silicon wafer p-type (boron doped), leading to different morphologies:

- The first process, thereafter denoted as \textbf{(VLS1)}, has a duration of 10 min with a pressure of 0.4 Tore (T). The nano-texture is shown in Fig.~\ref{fig:nanofil_pure}-(a): it consists of a one-layered texture of 7 $\mu$m-length NWs. Their average orientation is about 80$^{\circ}$ with respect to the horizontal plane (the smaller orientation angle of the few NWs at the forward plan, is due to that they have been broken during the cutting of the surface prior to SEM imaging).

- The second process, thereafter denoted as \textbf{(VLS2)}, has a duration of 60 min with a pressure of 0.1 T. The nano-texture is shown in Fig.~\ref{fig:nanofil_pure}-(b): it comprises a dense lower layer made of $\sim$ 2 $\mu$m short entangled NWs and an few NWs of 7 $\mu$m in length. The orientation is more irregular than for VLS1, and the average angle with horizontal is smaller. Some of them even have a "hook" shape at their top end.

The as-prepared SiNWs surface is hydrogen-terminated. However upon exposure to air, a thin native oxide layer is formed on the surface. This termination confers a superhydrophilic character to the surface with a contact angle close to zero \cite{Coffinier07}. Therefore due to its high roughness, all the tested liquids perfectly wet the surface, as predicted by the Wenzel equation (\ref{eq:wenzel}). In order to turn the surfaces super-omniphobic, they are coated with 1H,1H, 2H, 2H,- perfluorodecyltrichlorosilane (PFTS), a low surface-energy layer. The PFTS molecules are dissolved in hexane, and the surface is immersed in the solution for 6 hours at room temperature in a dry nitrogen purged glovebox. The resulting surface was rinsed with hexane, chloroform, iso-propanol and dried with a gentle stream of nitrogen. 

\section{Wetting properties}

\begin{figure}
\begin{center}
\includegraphics[scale=0.50]{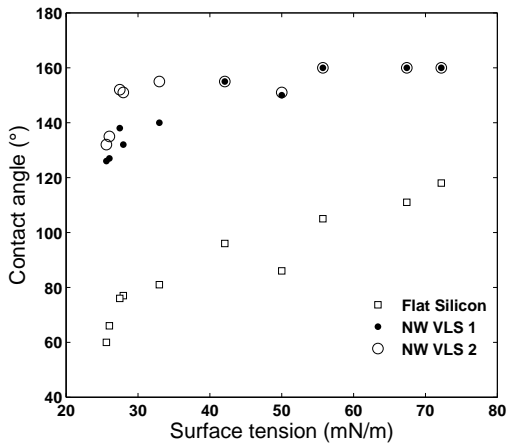}
\caption{Contact angle on flat silicon, NW VLS1 and NW VLS2 surfaces versus $\gamma$.}
\label{fig:contact_angles}
\end{center}
\end{figure}

We investigated liquids of surface tension varying from 25 to 72 mN/m. Liquids of lower $\gamma$, like n-hexane, spontaneously invade the NWs textures. For each liquid (see Table I), the CA and CAH on smooth silicon, NW-VLS1 and NW-VLS2 surfaces are measured with a goniometer (DSA100, Kruss GmbH Germany) and plotted in Figure \ref{fig:contact_angles}, versus surface tension. The contact angle $\theta_0$ of the drop on a smooth (flat) silicon surface is an appropriate measurement of the wettability of the surface for a given liquid. Indeed, the sole surface tension $\gamma$ between the liquid and its vapor cannot fully define this wettability, because the surface tension between the liquid and the solid $\gamma_{SL}$ is not \textit{a priori} known nor measurable. While $\theta_0$ on flat silicon decreases continuously as $\gamma$ decreases (with a CAH ranging from 20 to 30$^{\circ}$), the CA on the two NWs surfaces shows a plateau at about 160$^{\circ}$ -  and a small hysteresis (about 1$^{\circ}$) - for high $\gamma$ liquids, hence ensuring a super-omniphobic behavior. A sharp decrease of $\theta$ below 140$^{\circ}$ occurs at a threshold in $\gamma$. This threshold is about 33 mN/m for VLS1 and 26 mN/m for VLS2. However, despite the CA can fall down to 125$^{\circ}$, the CAH still remains weak (about a few degrees), which testifies that the drop remains in a Cassie-Baxter state. In most cases, VLS2 offers the best repelling performances.

\begin{table}
\caption{Physical properties of the investigated liquids.}
\begin{center}
\begin{ruledtabular}
\begin{tabular}{cccc}
Liquid & $\rho$ (g/cm$^3$) & $\nu$ (mm$^2$/s) & $\gamma$ (mN/m) \\
\hline
Water & 1.00 & 1.00 & 72.2  \\
Water+Glyc 50/50 & 1.126 & 4.93 & 67.4  \\
Water+Et 95/5 & 0.988 & 1.42 & 55.73  \\
CH$_2$I$_2$ & 3.32 & 2.26 & 50.0  \\
Water+Et 85/15 & 0.973 & 2.23 & 42.08  \\
Water+Et 70/30& 0.951 & 2.47 & 32.98  \\
Water+Et 50/50 & 0.910 & 2.20 & 27.96  \\
Hexadecane & 0.77 & 3.90 & 27.47  \\
Water+Et 38/62 & 0.882 & 1.98 & 26.02  \\
Water+Et 35/65 & 0.871 & 1.89 & 25.61 \\
\end{tabular}
\end{ruledtabular}
\end{center}
\label{tab:contact angle}
\end{table}

\section{Test of robustness : Drop impact experiments}

\begin{figure*}
\begin{center}
\includegraphics[scale=0.32]{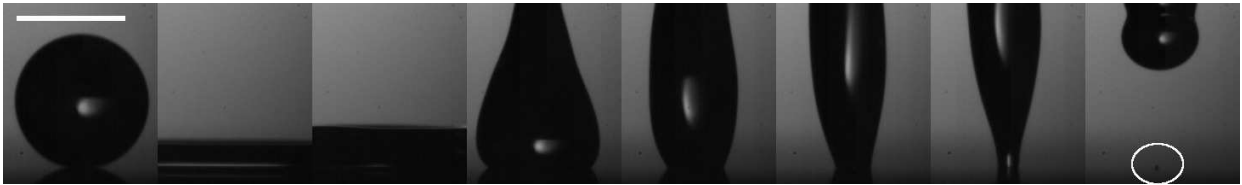}
\caption{A typical sequence of liquid impalement (here: di-iodomethane CH$_2$I$_2$ on VLS2 surface). A small drop remains, testifying the vicinity to threshold (see inside the white circle), as the drop bounces off the surface. The time between each snapshot is 2.67 ms. The bar in the first snapshot scales for 1 mm.}
\label{fig:setup_zoom}
\end{center}
\end{figure*}

\begin{figure}
\begin{center}
\includegraphics[scale=0.5]{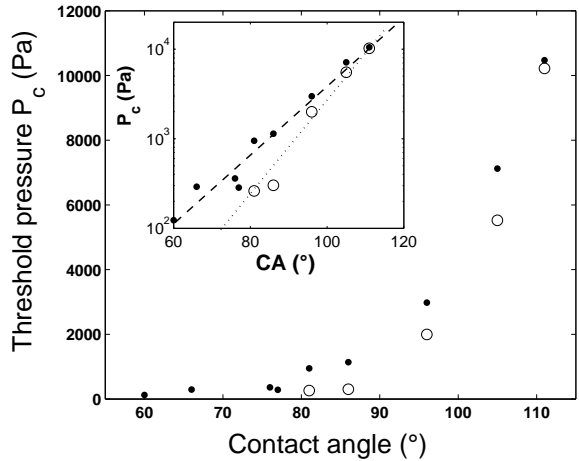}
\caption{Pressure threshold for liquid impalement versus the Young's contact angle $\theta_0$, that a drop of liquid adopts on an equivalent smooth surface. The open symbols stand for the VLS1 surface and the filled symbols stand for VLS2. Inset: Lin-Log plot.}
\label{fig:plot_rob}
\end{center}
\end{figure}

The set-up we used for drop impact has been previously presented in more details \cite{Brunet08,Faraday10}. We used a dripping faucet that releases a drop of liquid from a sub-millimetric nozzle. The drop detaches from the nozzle as the action of gravity overcomes the capillary retention forces. Therefore the diameter of the drop is determined by the capillary length of the liquid, and is well reproducible. Again, the properties of the liquids tested are given in Table I. The height of fall $h$ prescribes the velocity at impact: $V_0 = (2 g h)^{1/2}$, which can be up to 5.5 m/s. Using backlighting together with a high-speed camera (at a maximal rate of 4700 frames/s, with a resolution of 576$\times$576), the shape of the interface during the spreading and bouncing processes can be determined. The magnification allows for an accuracy of about 8 $\mu$m per pixel.

When released on a super-omniphobic surface, a drop of any tested liquid deforms and spreads like a pancake, then retracts and bounces off the surface, as previously observed in details \cite{Bartolo06,Quere_Reyssat06,Brunet08,Tsai09,Faraday10}. To determine $P_c$, we increase the height of fall $h$ (or impact velocity $U$) until the tiniest visible amount of  liquid remains at the impact location during the bouncing phase: this is observed in a typical sequence like in Figure \ref{fig:setup_zoom}. It is important to notice that the viscosity of the liquid $\nu$ influences a lot the dynamics of the spreading and bouncing phase: while the use of non-viscous liquid like water sometimes leads to an atomization of the main drop into many smaller droplets during the spreading phase (especially at high impact velocity), using large-$\nu$ liquids induces a large dissipation inside the texture. As a consequence, even if the drop has not impaled the texture down to the bottom (Wenzel state), the large viscous dissipation prevents the liquid from escaping the texture, as the bouncing height is lower than the size of the elongated drop. Therefore, we chose moderately viscous liquids, of the order of a few mm$^2$/s, in order to avoid the undesirable aforementioned effects.

It is also to be mentioned that we attempted experiments of impalement induced by droplet evaporation, in the same spirit as in \cite{Bartolo06,Tsai08}, in order to compare the obtained threshold with that measured with drop impact. However, due to the very low hysteresis (1 to 2 degrees), it was very hard to maintain a drop on our surfaces. The slightest departure from horizontality or the weakest wind around made the drop unexpectedly roll off the surface, which made it difficult to carry out such experiments.

Figure \ref{fig:plot_rob} shows the threshold pressure $P_c$ versus the contact-angle on flat silicon $\theta_0$. Whatever the liquid used, the threshold is always higher for the VLS2 surface as compared to VLS1 surface. The pressure $P_c$ goes to zero for CA $\theta_0$ smaller than 60$^{\circ}$, and seems to grow exponentially with $\theta_0$ (see inset). This plot is the confirmation of the high robustness of NWs surfaces against liquid impalement, even for low-$\gamma$ liquids. However, this robustness tends to deteriorate for highly wetting liquids, and there is a limit in surface tension (or in $\theta_0$) below which the liquid drop spontaneously gets impaled into the texture. It means that the capillary pressure itself is large enough to overcome the pressure threshold: $P_c \le \frac{2 \gamma}{R}$. Above this limit in $\theta_0$, the pressure threshold sharply increases: the growth can roughly be fitted by an exponential law, see Fig.~\ref{fig:plot_rob}.

From Fig.~\ref{fig:plot_rob}, it is clear that the VLS2 surface is more robust than the VLS1. This fact suggests that straight vertical NWs do not offer the best robustness, and that a rough, disordered lower layer should better improve the robustness. However, it is not possible to account quantitatively for the influence of surface roughness and the influence of NWs orientation, on the robustness, and to discriminate between the effect of these two factors.

This is clearly at odds to what was expected from previous measurements of $P_c$ with water on super-hydrophobic surfaces: a dense structure of tall and thin posts or NWs clearly have an advantage for a larger $P_c$ \cite{Bartolo06,Quere_Reyssat06,Brunet08}. We also conducted experiments with another type of surface which even more vertically-oriented NWs (the same as that denoted as 'P3' in ref. \cite{Brunet08}), and the robustness was even worse than for VLS1. We did not mention these experiments here because the aging of the surface, i.e. the formation of "bundles" of NWs similar to what was observed in carbon nanotubes (see Fig. 6 of \cite{Lau03}) prevented reproducible experiments). Therefore, our results reveal that it is crucial to have diagonally grown NWs in order to get a good robustness for low-$\gamma$ liquids.

\section{Conclusions}

In conclusion, we presented the first quantitative measurements of robustness of a super-omniphobic surface. We measured the pressure threshold for liquid impalement on two different NWs surfaces for various liquids of different surface tension, which allowed to test the influence of the Young's contact angle $\theta_0$ that a liquid drop adopts on a smooth surface of the same surface energy. It turns out that $\theta_0$ plays a crucial role in the robustness: for a surface of given structure and roughness, there exists a threshold value for $\theta_0$ below which one observes spontaneous impalement (without drop impact) of liquid. Above this threshold in $\theta_0$, the limit pressure sharply increases with $\theta_0$.

The surfaces made of a carpet of nano-wires, oriented with a finite angle with respect to the vertical, offered a liquid-repellent character with very good robustness over a large range of $\gamma$, and a nano-scale equivalent of the re-entrant structures proposed in previous studies \cite{Tuteja07,Tuteja08}. However, it is not yet clear whether it is the average orientation or the relative disorder of the NWs that contributes to a better robustness. Future studies will be focused on elucidating these points.

\section*{Acknowledgements} 

We are indebted to the BioMems team of IEMN for the technical support that was required to carry out this experimental study. The European Community's Seventh Framework Programme (FP7/2007Ð2013) under grant agreement no. 227243, the Centre National de la Recherche Scientifique (CNRS) and the Nord-Pas-de Calais region are gratefully acknowledged for financial support.

\end{document}